\documentclass[floafix,prd,showpacs,showkeys,preprintnumbers,nofootinbib,superscriptaddress,11pt]{revtex4-1}
\usepackage[utf8]{inputenc}
\usepackage[sort&compress]{natbib}
\usepackage{ulem}
\usepackage{bm}
\usepackage{times}
\usepackage{amssymb,amsbsy,amsmath,amsfonts}
\usepackage{graphicx}
\usepackage{epstopdf}
\usepackage{float}
\usepackage{color}
\usepackage{morefloats}
\usepackage{rotating}
\usepackage{srcltx}
\usepackage{slashed}
\usepackage{subfigure}
\usepackage{multirow}
\usepackage{verbatim}
\usepackage{hyperref}
\usepackage{overpic}
\usepackage{tabularx}

\begin{document}

\title{Production of $D^*_{s0}(2317)$ and $D_{s1}(2460)$ in  $B$ decays  as  $D^{(*)}K$ and $D^{(*)}_s\eta$ molecules}

\author{Ming-Zhu Liu}
\affiliation{School of Space and Environment, Beihang University, Beijing 102206, China}
\affiliation{School of Physics, Beihang University, Beijing 102206, China}

\author{Xi-Zhe Ling}
\affiliation{Institute of High Energy Physics,
Chinese Academy of Sciences,
Beijing 100049, China}

\author{Li-Sheng Geng}~\email{lisheng.geng@buaa.edu.cn}
\affiliation{Peng Huanwu Collaborative Center for Research and Education, Beihang University, Beijing 100191, China}
\affiliation{School of Physics, Beihang University, Beijing 102206, China}
\affiliation{Beijing Key Laboratory of Advanced Nuclear Materials and Physics, Beihang University, Beijing, 102206, China}
\affiliation{School of Physics and Microelectronics, Zhengzhou University, Zhengzhou, Henan 450001, China}

\author{En Wang}~\email{wangen@zzu.edu.cn}
\affiliation{School of Physics and Microelectronics, Zhengzhou University, Zhengzhou, Henan 450001, China}

\author{Ju-Jun Xie}~\email{xiejujun@impcas.ac.cn}
\affiliation{Institute of Modern Physics, Chinese Academy of Sciences, Lanzhou 730000, China}
\affiliation{School of Nuclear Science and Technology, University of Chinese Academy of Sciences, Beijing 101408, China}
\affiliation{School of Physics and Microelectronics, Zhengzhou University, Zhengzhou, Henan 450001, China}
\affiliation{Lanzhou Center for Theoretical Physics, Key Laboratory of Theoretical Physics of Gansu Province, Lanzhou University, Lanzhou, Gansu 730000, China}

\begin{abstract}
The molecular nature of $D_{s0}^{\ast}(2317)$ and $D_{s1}(2460)$ have been extensively studied from the perspective of their masses, decay properties, and production rates. 
In this work,  we study the weak decays of $B \to \bar{D}^{(\ast)}D_{s0}^{*}(2317)$ and  $B \to \bar{D}^{(\ast)}D_{s1}(2460)$ by invoking  triangle diagrams where the $B$ meson first decays weakly into $\bar{D}^{(\ast)}D_{s}^{(\ast)}$ and $J/\psi K$($\eta_{c}K$), and then the $D_{s0}^{\ast}(2317)$ and $D_{s1}(2460)$  are  dynamically generated by the final-state interactions of $D_{s}^{(\ast)}\eta$ and $D^{(\ast)}K$ via exchanges of $\eta$ and $D^{(\ast)}$  mesons.  The obtained absolute branching fractions of Br$[B \to \bar{D}^{(\ast)}D_{s0}^{*}(2317)]$ are in reasonable agreement with the experimental data, while the branching fractions of Br$[B \to \bar{D}^{(\ast)}D_{s1}(2460)]$ are smaller than the  experimental central values by almost a factor of two to three. We tentatively attribute such a discrepancy to either  reaction mechanisms missing in the present work or the likely existence of a relatively larger $c\bar{s}$ component in the $D_{s1}(2460)$ wave function.

\end{abstract}


\maketitle

\section{Introduction}
In 2003, the BaBar Collaboration discovered a quite narrow state near 2.32 GeV in the inclusive $D_{s}^{+}\pi^{0}$ invariant  mass distribution~\cite{BaBar:2003oey}, named as $D_{s0}^{\ast}(2317)$, which was subsequently confirmed  by the CLEO~\cite{CLEO:2003ggt} and Belle Collaborations~\cite{Belle:2003guh}. Taken  as a $c\bar{s}$ state  with the quantum number of $I(J^P) =  0 (0^+)$, its mass is  lower by 160 MeV than the 
prediction of the Godfrey-Isgur (GI) quark model~\cite{Godfrey:1985xj}. Such a large deviation  has also appeared  within the lattice QCD simulations~\cite{Moir:2013ub,Cheung:2016bym}.   To explain the discrepancy, many different  interpretations of the $D_{s0}^*(2317)$ have been proposed, such as a $P$-wave $c\bar{s}$ excited state~\cite{Cahn:2003cw,Godfrey:2003kg,Colangelo:2003vg}, a compact tetraquark state~\cite{Maiani:2004vq}, or a hadronic molecule~\cite{Kolomeitsev:2003ac,vanBeveren:2003kd,Barnes:2003dj,Chen:2004dy,Guo:2006rp}. Among them, the hadronic molecular interpretation has attracted considerable attention. 

In Refs.~\cite{Guo:2006fu,Gamermann:2006nm}, the authors interpreted $D_{s0}^*(2317)$ as a hadronic molecule generated by the $DK$ and $D_{s}\eta$ coupled-channel interactions in the chiral unitary approach, which is also supported by many other studies~\cite{Altenbuchinger:2013vwa,Yao:2015qia,Guo:2015dha,Du:2017ttu}. The $DK$ coupled-channel interactions~\cite{Liu:2012zya,Lang:2014yfa,Cheung:2020mql} have been simulated on the lattice, and a bound state below the $DK$ mass threshold  is found, which can be identified as $D_{s0}^{*}(2317)$. In addition, a $D^{\ast}K$ molecule as the partner of $D_{s0}^*(2317)$ is predicted via the heavy quark spin symmetry (HQSS), and it can be identified as  $D_{s1}(2460)$~\cite{Altenbuchinger:2013vwa,Guo:2006rp,Bali:2017pdv,Hu:2020mxp},  discovered by  the
CLEO Collaboration in the $D_{s}^*\pi$ mass distribution~\cite{CLEO:2003ggt} and confirmed by the Belle Collaboration~\cite{Belle:2003guh}. Up to now, only the upper limits for the  widths of $D_{s0}^*(2317)$ and $D_{s1}(2460)$ are known, i.e., $\Gamma_{D_{s0}^*(2317)}<3.7$ MeV and $\Gamma_{D_{s1}(2460)}<3.5$ MeV~\cite{ParticleDataGroup:2020ssz}. In the molecular picture, Faessler et al. took the effective Lagrangian approach to estimate the the dominant partial decay widths of $D_{s0}^{*}(2317) \to D_{s}\pi$ and $D_{s1}(2460) \to D_{s}^*\pi$ to be 80 keV and 50$\sim$79 keV~\cite{Faessler:2007gv,Faessler:2007us}. Very recently, an effective field theory study estimated their partial decay widths to be 120 keV and 102 keV~\cite{Fu:2021wde}, respectively.

Recently, we proposed a novel approach to verify the molecular nature of exotic states from the existence of relevant three-hadron molecules (see Refs.~\cite{Wu:2021dwy,Wu:2022ftm} for reviews). The molecular nature of $D_{s0}^*(2317)$ can be verified by searching for  the three-body molecule $DDK$, where the $DK$ interaction is  determined by reproducing the mass of $D_{s0}^*(2317)$ and plays a dominant role in forming the $DDK$ molecule~\cite{Wu:2019vsy,Huang:2019qmw}. 
In Ref.~\cite{SanchezSanchez:2017xtl}, assuming $D_{s0}^*(2317)$ as a $DK$ molecule,  we employed the one-kaon-exchange potential and predicted  the existence of
a $DD_{s0}^*(2317)$ molecule, whose mass and quantum numbers are consistent with those of the $DDK$ molecule. Moreover, we have investigated the $\bar{D}DK$ system~\cite{Wu:2020job}, and it was found that the $\bar{D}D_{s0}^*(2317)$ configuration accounts for  about $87\%$ of the $\bar{D}DK$ configuration, which indicates that  the $DK$ interaction plays the most important role in forming the $\bar{D}DK$ molecule as well~\cite{Wei:2022jgc}. If the $\bar{D}DK$ molecule is discovered by experiments, it will also verify the molecular nature of $D_{s0}^*(2317)$. It should be noted that although the $DK$ molecular interpretation is the most favorable, the $c\bar{s}$ component is found to play a non-negligible role in describing the mass of $D_{s0}^{*}(2317)$ in the unquenched quark models~\cite{Dai:2003yg,Simonov:2004ar,Ortega:2016mms,Luo:2021dvj,Hao:2022vwt}.
In a recent work~\cite{Yang:2021tvc}, by fitting to the lattice QCD finite volume spectra, Yang et al. found that the $c\bar{s}$ component accounts for about $32\%$ of the wave function of $D_{s0}^*(2317)$, while the $c\bar{s}$ component accounts for more than half of the $D_{s1}(2460)$ wave function,  which is consistent with a number of earlier studies~\cite{MartinezTorres:2014kpc,Albaladejo:2018mhb,Tan:2021bvl}. 

The production  of $D_{s0}^*(2317)$ in the molecular picture has also been  extensively investigated. In Ref.~\cite{ExHIC:2011say}, assuming  $D_{s0}^{\ast}(2317)$ as either  a conventional  $c\bar{s}$ state, a compact multiquark state or a hadronic molecule,  Cho {\it et al}. adopted the coalescence model and statistical model to estimate the corresponding yield of $D_{s0}^{\ast}(2317)$ in heavy ion collisions, which would help probe its nature  in future experiments. On the other hand, the production of $D_{s0}^*(2317)$ in the weak decays of $B$ and $B_{s}$ mesons  also provides a very good platform to study the meson-meson interactions and the nature of $D_{s0}^*(2317)$.   In Ref.~\cite{Albaladejo:2016hae}, Miguel {\it et al}. investigated the nature of $D_{s0}^{\ast}(2317)$ by extracting the $DK$ interaction via the $DK$ invariant mass distributions of the processes $B^{+}\to \bar{D}^{0}D^{0}K^{+}$,  $B^{0}\to {D}^{-}D^{0}K^{+}$, and $B^0_{s}\to \pi^{+}\bar{D}^{0}K^{-}$. In Ref.~\cite{Navarra:2015iea}, Navarra {\it et al}. investigated the molecular nature of $D_{s0}^{\ast}(2317)$ in the semileptonic $B^0_{s}$ and $B$ decays  taking into account the $DK$ and $D_{s}\eta$ rescattering. 

On the experimental side, the $D_{s0}^{\ast}(2317)$ and $D_{s1}(2460)$ have been found in the weak decays of $B\to \bar{D}^{(\ast)}D_{s0}^*(2317)$ and $B \to \bar{D}^{(\ast)}D_{s1}(2460)$, and their branching fractions can be found in Ref.~\cite{ParticleDataGroup:2020ssz}. 
In Ref.~\cite{Cheng:2006dm}, Cheng {\it et al}. employed  the covariant light-front quark model to study the weak decays of $B\to \bar{D}^{(\ast)}D_{s0}^*(2317)$ and $B\to \bar{D}^{(\ast)}D_{s1}(2460)$ using the factorization approach, where  $D_{s0}^{\ast}(2317)$ and $D_{s1}(2460)$ are treated as  $P$-wave $c\bar{s}$ states. Later, Segovia {\it et al}. adopted a similar approach  to study the decays $B\to \bar{D}^{(\ast)}D_{s0}^*(2317)$ and $B\to \bar{D}^{(\ast)}D_{s1}(2460)$~\cite{Segovia:2012yh}. Recently, Zhang {\it et al}. calculated the decay  $B\to \bar{D}^{(\ast)}D_{s0}^*(2317)$ in the pQCD approach~\cite{Zhang:2021bcr}.    In addition, the production rates of $D_{s1}^*(2317)$ and $D_{s1}(2460)$ in the semileptonic decays  $B_{s}\to D_{s0}^*(2317)(D_{s1}^*(2460))l\bar{v}$~\cite{Huang:2004et} and in the nonleptonic decays  $\Lambda_{b}\to \Lambda_{c}D_{s0}^*(2317)(D_{s1}(2460))$~\cite{Datta:2003yk} have been predicted.

Assuming  $D_{s0}^{\ast}(2317)$ and $D_{s1}(2460)$ as  $DK$ and $D^*K$  molecules, Faessler {\it et al}. calculated the branching ratios of $B\to \bar{D}^{(\ast)}D_{s0}^*(2317)$ and $B \to \bar{D}^{(\ast)}D_{s1}(2460)$ in the naive factorization approach~\cite{Faessler:2007cu}, where the couplings $f_{D_{s0}^*}$ and $f_{D_{s1}}$ are estimated in the molecular picture, different from Refs.~\cite{Cheng:2006dm,Segovia:2012yh}. 
In the present work, we will revisit  the $B\to \bar{D}^{(\ast)}D_{s0}^*(2317)$ and $B \to \bar{D}^{(\ast)}D_{s1}(2460)$ decays in the triangle mechanism, where $D_{s0}^*(2317)$ and  $D_{s1}(2460)$ are dynamically generated by the coupled-channels $D^{(\ast)}K$ and $D_s^{(\ast)}\eta$. We note that a similar approach has  earlier been employed to study $a_{0}(980)$ generated by the coupled-channels $\pi\eta$ and $K\bar{K}$ in the process $D_{s}\to\pi\pi\eta$~\cite{Ling:2021qzl}, where the theoretical results are found in good agreement with the experimental data. 

This work is organized as follows. We briefly introduce the triangle mechanism for the decays of $B\to \bar{D}^{(\ast)}D_{s0}^*(2317)$ and $B\to \bar{D}^{(\ast)}D_{s1}(2460)$ and the effective Lagrangian approach in Sec.~II. Results and discussions are given in Sec.~III, followed by a short summary in the last section.

\section{Theoretical formalism}

The mesonic weak transition form factors and decay constants are the two main ingredients in the study of hadronic weak decays of mesons, which are less certain for  $P$-wave charmed mesons than for $S$-wave charmed mesons. Here, we adopt the triangle mechanism to study the weak decays of $B\to \bar{D}^{(\ast)}D_{s0}^*(2317)$ and $B\to \bar{D}^{(\ast)}D_{s1}(2460)$, where the form factors and decay constants of $S$-wave mesons are stringently constrained by experiments. This way, we can largely reduce the theoretical uncertainties.  In the following, we explain in detail the triangle mechanism accounting for the weak decays of $B\to \bar{D}^{(\ast)}D_{s0}^*(2317)$ and $B\to \bar{D}^{(\ast)}D_{s1}(2460)$.

\subsection{Triangle diagrams}

\begin{figure}[htbp]
\begin{center}
\subfigure[]
{
\begin{minipage}[t]{0.47\linewidth}
\begin{center}
\begin{overpic}[scale=.68]{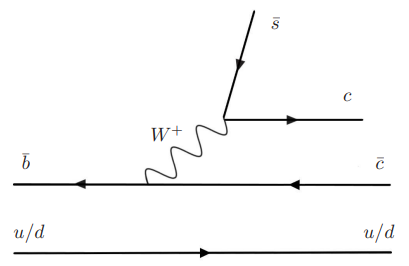}
\end{overpic}
\end{center}
\end{minipage}
}
\subfigure[]
{
\begin{minipage}[t]{0.47\linewidth}
\begin{center}
\begin{overpic}[scale=.68]{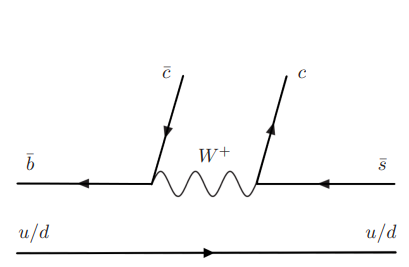}
\end{overpic}
\end{center}
\end{minipage}
}
  \caption{(a) External $W$-emission  for $B^{+(0)}\to D_s^{(\ast)+}\bar{D}^{*0}(D^{*-})$ and (b) internal $W$-conversion  for $B^{+(0)}\to J/\psi/\eta_{c} K^{+(0)}$.}
  \label{quark}
\end{center}
\end{figure}

At the quark level, the decays of $B^{+(0)}\to D_{s}^{(\ast)+} \bar{D}^{(\ast)0}(D^{(\ast)-})$ and  $B^{+(0)}\to J/\psi(\eta_{c}) K^{+(0)}$ can  proceed via the external $W$-emission and  the internal $W$-conversion mechanisms  as shown in Fig.~\ref{quark}(a) and (b), respectively.  Referring to the Review of Particle Physics(RPP)~\cite{ParticleDataGroup:2020ssz} , the absolute branching fractions of the processes  $B^{+(0)}\to D_{s}^{(\ast)+} \bar{D}^{(\ast)0}(D^{(\ast)-})$ and  $B^{+(0)}\to J/\psi(\eta_{c}) K^{+(0)}$ are tabulated in Table~\ref{exp}, which 
follows the topological classification of  weak decays where the strength of the external $W$-emission mechanism is larger than that of  the internal $W$-conversion mechanism~\cite{Chau:1982da,Chau:1987tk,Molina:2019udw}.

\begin{table}[htbp]
\centering
\caption{Branching ratios ($10^{-3}$) of  $B^{+(0)}\to D_{s}^{(\ast)+} \bar{D}^{(\ast)0}(D^{(\ast)-})$ and  $B^{+(0)}\to J/\psi(\eta_{c}) K^{+(0)}$. \label{exp}
}
\label{ratios}
\begin{tabular}{c c c c c c c c c}
  \hline \hline
    Decay mode    &~~~~ RPP.~\cite{ParticleDataGroup:2020ssz}  & Decay mode    &~~~~ RPP.~\cite{ParticleDataGroup:2020ssz}  
         \\ \hline 
        $B^{+} \to \bar{D}^{0}D_{s}^{+}$    & ~~~~ $9.0\pm 0.9$    &         $\bar{B}^{0} \to {D}^{-}D_{s}^{+}$    & ~~~~ $7.2\pm 0.8$  
        \\
                $B^{+} \to \bar{D}^{0}D_{s}^{\ast+} $ & ~~~~ $7.6\pm 1.6$ &         $\bar{B}^{0} \to {D}^{-}D_{s}^{\ast+}$    & ~~~~ $7.4\pm 1.6$  
         \\
                $B^{+} \to \bar{D}^{\ast0}D_{s}^{+} $ & ~~~~ $8.2\pm 1.7$ &         $\bar{B}^{0} \to {D}^{\ast-}D_{s}^{+}$    & ~~~~ $8.0\pm 1.1$  
         \\  
                         $B^{+} \to \bar{D}^{\ast0}D_{s}^{\ast+} $ & ~~~~ $17.1\pm 2.4$&         $\bar{B}^{0} \to {D}^{\ast-}D_{s}^{\ast+}$    & ~~~~ $17.7\pm 1.4$  
         \\  
                               $B^{+} \to J/\psi K^{+} $ & ~~~~ $1.010\pm 0.029$ &                            $\bar{B}^0 \to J/\psi K^{0} $ & ~~~~ $0.873\pm 0.032$
         \\  
                             $B^{+} \to \eta_{c} K^{+} $ & ~~~~ $1.09\pm 0.09$ &      $\bar{B}^0 \to \eta_{c} K^{0} $ & ~~~~ $0.79\pm 0.12$
         \\  
  \hline \hline
\end{tabular}
\end{table}

\begin{figure}[htbp]
\begin{center}
\subfigure[]
{
\begin{minipage}[t]{0.45\linewidth}
\begin{center}
\begin{overpic}[scale=.58]{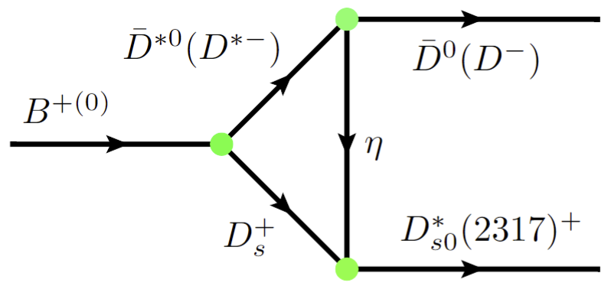}
\end{overpic}
\end{center}
\end{minipage}
}
\subfigure[]
{
\begin{minipage}[t]{0.45\linewidth}
\begin{center}
\begin{overpic}[scale=.84]{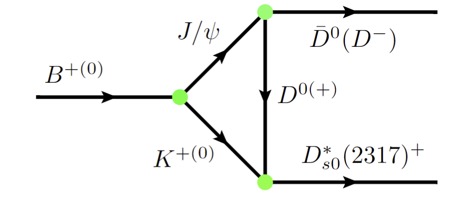}
\end{overpic}
\end{center}
\end{minipage}
}
\subfigure[]
{
\begin{minipage}[t]{0.45\linewidth}
\begin{center}
\begin{overpic}[scale=.86]{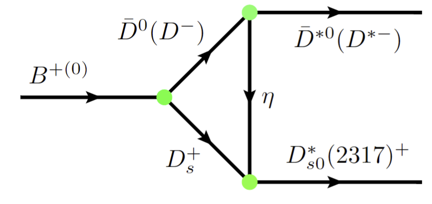}
\end{overpic}
\end{center}
\end{minipage}
}
\subfigure[]
{
\begin{minipage}[t]{0.45\linewidth}
\begin{center}
\begin{overpic}[scale=.65]{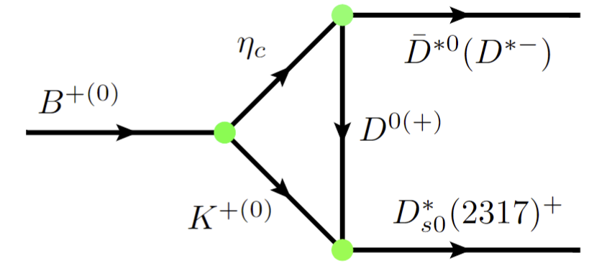}
\end{overpic}
\end{center}
\end{minipage}
}
\caption{Triangle diagrams accounting for the four $B$ decays: (a) $B^{+(0)}\to D_s^+\bar{D}^{*0}(D^{*-})\to D_{s0}^*(2317)\bar{D}^0(D^-)$, (b) $B^{+(0)}\to J/\psi K^{+(0)}\to D_{s0}^*(2317)\bar{D}^0(D^-)$, (c) $B^{+(0)}\to D_s^+\bar{D}^{0}(D^{-})\to D_{s0}^*(2317)\bar{D}^{\ast0}(D^{\ast-})$ and (d) $B^{+(0)}\to \eta_{c} K^{+(0)}\to D_{s0}^*(2317)\bar{D}^{\ast0}(D^{\ast-})$.   }
\label{triangle2317}
\end{center}
\end{figure}

\begin{figure}[htbp]
\begin{center}
\subfigure[]
{
\begin{minipage}[t]{0.45\linewidth}
\begin{center}
\begin{overpic}[scale=.9]{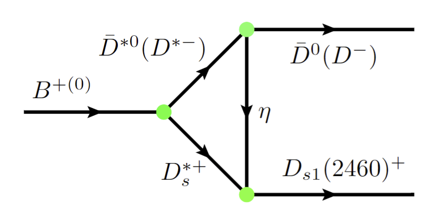}
\end{overpic}
\end{center}
\end{minipage}
}
\subfigure[]
{
\begin{minipage}[t]{0.45\linewidth}
\begin{center}
\begin{overpic}[scale=.9]{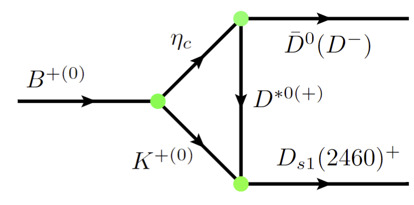}
\end{overpic}
\end{center}
\end{minipage}
}
\caption{Triangle diagrams accounting for the two $B$ decays: (a) $B^{+(0)}\to D_s^{\ast+}\bar{D}^{*0}(D^{*-})\to D_{s1} (2460)^+\bar{D}^0(D^-)$ and (b) $B^{+(0)}\to \eta_{c} K^{+(0)}\to D_{s1} (2460)^+\bar{D}^0(D^-)$. }
\label{triangle24601}
\end{center}
\end{figure}

\begin{figure}[!h]
\begin{center}
\subfigure[]
{
\begin{minipage}[t]{0.45\linewidth}
\begin{center}
\begin{overpic}[scale=.67]{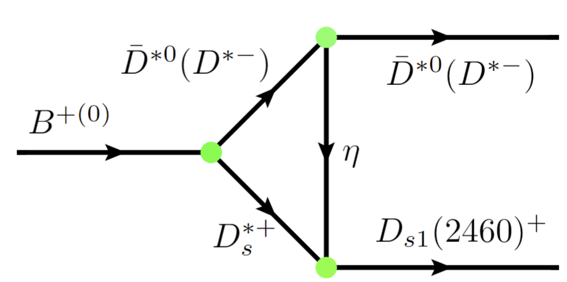}
\end{overpic}
\end{center}
\end{minipage}
}
\subfigure[]
{
\begin{minipage}[t]{0.45\linewidth}
\begin{center}
\begin{overpic}[scale=.8]{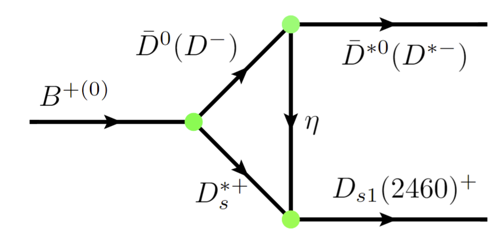}
\end{overpic}
\end{center}
\end{minipage}
}
\subfigure[]
{
\begin{minipage}[t]{0.45\linewidth}
\begin{center}
\begin{overpic}[scale=.8]{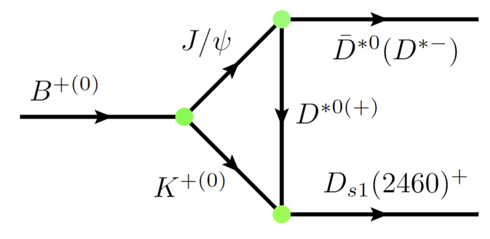}
\end{overpic}
\end{center}
\end{minipage}
}  
\subfigure[]
{
\begin{minipage}[t]{0.45\linewidth}
\begin{center}
\begin{overpic}[scale=.8]{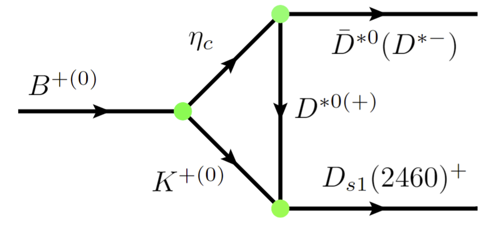}
\end{overpic}
\end{center}
\end{minipage}
}
\caption{Triangle diagrams accounting for the four $B$ decays: (a) $B^{+(0)}\to D_s^{\ast+}\bar{D}^{*0}(D^{*-})\to D_{s1} (2460)^+\bar{D}^{\ast0}(D^{\ast-})$, (b) $B^{+(0)}\to D_s^{\ast+}\bar{D}^{0}(D^{-})\to D_{s1} (2460)^+\bar{D}^{\ast0}(D^{\ast-})$, (c) $B^{+(0)}\to J/\psi K^{+(0)}\to D_{s1} (2460)^+\bar{D}^{\ast0}(D^{\ast-})$, and  (d) $B^{+(0)}\to \eta_{c} K^{+(0)}\to D_{s1} (2460)^+\bar{D}^{\ast0}(D^{\ast-})$   }
\label{triangle24602}
\end{center}
\end{figure}

Taking into account the scattering vertices of $\bar{D}^{\ast}\to \bar{D}\eta$,   $J/\psi \to \bar{D}D$, $\bar{D}\to \bar{D}^{\ast}\eta$ and  $\eta_{c} \to \bar{D}^{\ast}D$,   the $D_{s0}^*(2317)$ state can be dynamically generated by the $DK$ and $D_{s}\eta$ coupled-channel interactions.  We illustrate the decays of $B^{+}\to \bar{D}^{(\ast)0}D_{s0}^*(2317)^{+}$ and $B^{0}\to D^{(\ast)-}D_{s0}^*(2317)^{+}$ at the hadronic level via the triangle diagrams  shown in Fig.~\ref{triangle2317}. Similarly, we depict the triangle diagrams of the decays of $B^{+(0)}\to \bar{D}^{0}(D^{-})D_{s1}(2460)^{+}$ in Fig.~\ref{triangle24601}, and $B^{+(0)}\to \bar{D}^{\ast0}(D^{\ast-})D_{s1}(2460)^{+}$ in Fig.~\ref{triangle24602}.

\subsection{Effective Lagrangians}
 
To compute the contributions of the triangle diagrams shown in Figs.~\ref{triangle2317}, \ref{triangle24601}, and \ref{triangle24602}, we introduce the effective Lagrangians. The effective Hamiltonian describing the weak decays of $B^{+(0)}\to  D_{s}^{(\ast)+} \bar{D}^{0(\ast)}(D^{-(\ast)})$ and  $B^{+(0)}\to J/\psi(\eta_{c}) K^{+(0)}$ has the following form
\begin{equation}
\mathcal{H}_{eff}=\frac{G_F}{\sqrt{2}}V_{cb}V_{cs}[c_1^{eff}\mathcal{O}_{1}+c_2^{eff}\mathcal{O}_{2}]+h.c.,
\end{equation}
where $G_F $ is the Fermi constant, $V_{bc}$ and $V_{cs}$ are the  Cabibbo-Kobayashi-Maskawa(CKM) matrix elements, $c_{1,2}^{eff}$ are the effective Wilson coefficients, and $\mathcal{O}_{1}$ and $\mathcal{O}_{2}$ are the four-fermion operators of  $(s\bar{c})_{V-A}(c\bar{b})_{V-A}$ and $(\bar{c}c)_{V-A}(s\bar{b})_{V-A}$ with $(q\bar{q})_{V-A}$ standing for $q\gamma_\mu(1-\gamma_5)\bar{q}$~\cite{Buras:1998raa,Geng:2017esc,Han:2021azw}.

The effective  Lagrangians accounting for the interactions between the charmonium states ($J/\psi$, $\eta_{c}$) and a pair of charmed mesons read~\cite{Xiao:2019mvs,Wu:2021ezz}  
\begin{eqnarray}
\mathcal{L}_{\psi DD}&=&ig_{\psi DD} \psi_{\mu}(\partial^{\mu}D\bar{D}-D\partial^{\mu}\bar{D}), \\ \nonumber 
\mathcal{L}_{\psi DD^{\ast}}&=&-g_{\psi DD^{\ast}} \epsilon^{\alpha \beta \mu \nu}\partial_{\alpha} \psi_{\beta}(\partial_{\mu} D_{\nu}^{\ast}\bar{D}+ D \partial_{\mu}\bar{D}^{\ast}_{\nu}), 
\\ \nonumber 
\mathcal{L}_{\psi D^{\ast}D^{\ast}}&=&-ig_{\psi D^{\ast}D^{\ast}} [\psi^{\mu}(\partial_{\mu}D^{\ast}_{\nu}\bar{D}^{\ast\nu}-D^{\ast \nu}\partial_{\mu}\bar{D}^{\ast}_{\nu}) +(\partial_{\mu}\psi_{\nu}D^{\ast\nu}-\psi_{\nu}\partial_{\mu}D^{\ast\nu})\bar{D}^{\ast\mu} \\  \nonumber
&&+ {D}^{\ast\mu} (\psi^{\nu}\partial_{\mu}\bar{D}^{\ast}_{\nu}-\partial_{\mu}\psi_{\nu}\bar{D}^{\ast\nu})],  \\ \nonumber
\mathcal{L}_{\eta_{c} D^{\ast}D}&=&ig_{\eta_{c} D^{\ast}D}[D^{\ast\mu}(\partial_{\mu}\eta_{c}\bar{D}-\eta_{c}\partial_{\mu}\bar{D})-(\partial_{\mu}\eta_{c}D-\eta_{c}\partial_{\mu}D)\bar{D}^{\ast\mu}],   \\ \nonumber
\mathcal{L}_{\eta_{c} D^{\ast}D^{\ast}}&=&-g_{\eta_{c} D^{\ast}D^{\ast}}\varepsilon^{\mu\nu\alpha\beta}\partial_{\mu}D^{\ast}_{\nu}{\partial}_{\alpha} \bar{D}^{\ast}_{\beta}\eta_{c}, 
\end{eqnarray}
where $g_{\psi DD}$,  $g_{\psi D^{\ast}D}$,  $g_{\psi D^{\ast}D^\ast}$, $g_{\eta_{c} D^{\ast}D}$, and $g_{\eta_{c} D^{\ast}D^{\ast}}$ are the couplings of the charmonium mesons  to the charmed mesons. The coupling constants   are determined as follows:  $g_{\psi DD}=g_{\psi D^*D^*}=m_{\psi}/f_{\psi}$, $g_{\psi D^*D}=\frac{2}{m_{D}}g_{\psi DD}$~\cite{Oh:2000qr,Wu:2021ezz}, $g_{\eta_{c} D^* D}=\frac{m_{D}}{2}g_{\eta_{c} D^*D^*}=g_{2}\sqrt{m_{\eta_c}}m_{D}$,  and $g_{2}=2.36$~GeV$^{-3/2}$~\cite{Lin:2017mtz}.

The effective Lagrangian  describing the interaction between the charmed mesons ($D$ and $D^*
$) and $\eta$ are written as~\cite{Wu:2021cyc}  
\begin{eqnarray}
\mathcal{L}_{D D^{\ast} \eta}&=& i g_{D D^{\ast} \eta} (D_{\mu}^* \partial^{\mu} \eta  \bar{D}-D \partial^{\mu}\eta \bar{D}^{\ast}_{\mu}),   \\ \nonumber
\mathcal{L}_{D^{\ast} D^{\ast} \eta}&=& - g_{D^{\ast} D^{\ast} \eta} \varepsilon_{\mu\nu\alpha\beta} \partial^{\mu}D^{\ast\nu} {\partial}^{\alpha} \bar{D}^{\ast\beta}\eta ,  
\end{eqnarray}
where $g_{D D^{\ast} \eta}$ and $g_{D^{\ast} D^{\ast} \eta}$ are the couplings between charmed mesons and light mesons. For the couplings between the charmed mesons and $\eta$, $g_{D^{*0}D^0\eta}=g_{D^{*-}D^-\eta}=\frac{g_{D^{*0}D^0\pi^{0}}}{\sqrt{3}}$  are derived by the SU(3)-flavor symmetry, and the coupling  $g_{D^{*0}D^0\pi^{0}}=11.7$ is obtained from the decay width of $D^{*0}\to D^0\pi^0$ \cite{ParticleDataGroup:2020ssz}. The coupling of $g_{\bar{D}^{*0}\bar{D}^{*0}\eta}$ is obtained by the relationship  $g_{\bar{D}^{*0}\bar{D}^{*0}\eta}=g_{D^{*0}D^0\eta}/m_{D}$~\cite{Wu:2021cyc}.

Assuming that $D_{s0}^*(2317)$ and $D_{s1}(2460)$ are dynamically generated by the $S$-wave $DK$-$D_{s}\eta$ and $D^*K$-$D_{s}^*\eta$ coupled-channel interactions, respectively,  the relevant Lagrangians can be written as~\cite{Faessler:2007gv,Faessler:2007us} 
\begin{eqnarray}
\mathcal{L}_{D_{s0}^* D K} & = & g_{D_{s 0}^* D K} D_{s 0}^* D K, \\  \nonumber
\mathcal{L}_{D_{s 0}^* D_{s} \eta} & = & g_{D_{s 0}^* D_{s} \eta} D_{s 0}^* D_{s} \eta,  \\ \nonumber
\mathcal{L}_{D_{s1}D^{\ast}K}&=&g_{D_{s1}D^{\ast}K}D_{s1}^{\mu}D^{\ast}_{\mu}K,  \\ \nonumber
\mathcal{L}_{D_{s1}D_{s}^{\ast}\eta}&=&g_{D_{s1}D_{s}^{\ast}\eta}D_{s1}^{\mu}D_{s\mu}^{\ast}\eta,
\end{eqnarray}
where  $g_{D_{s0}^* DK}$ and $g_{D_{s0}^*D_{s}\eta}$ represent the  couplings of $D_{s0}^*(2317)$ to $DK$ and $D_{s}\eta$, and $g_{D_{s1}D^{\ast}K}$ and $g_{D_{s1}D_{s}^{\ast}\eta}$ represent the couplings of $D_{s1}(2460)$ to $D^*K$ and $D_{s}^*\eta$. The values of $g_{D_{s0}^{\ast}DK}$ and $g_{D_{s0}^{\ast}D_{s}\eta}$ are determined from the residues of $D_{s0}^{\ast}(2317)$ on the complex plane, where it is treated as  a molecule dynamically  generated by the  $DK$ and $D_{s}\eta$ coupled-channel interactions. In this work, we take $g_{D_{s0}^* DK}=9.4$ GeV and $g_{D_{s0}^* D_{s}\eta}=7.4$ GeV given in the effective field theory approach~\cite{Fu:2021wde}, in agreement with those obtained in Ref.~\cite{Gamermann:2006nm}. The $D_{s1}(2460)$ is regarded as the HQSS partner of $D_{s0}^{\ast}(2317)$, which is dynamically generated  by the $D^*K$ and $D_{s}^*\eta$ coupled-channel interactions. The couplings of $g_{D_{s1} D^*K}=10.1$ GeV and $g_{D_{s} D_{s}^*\eta}=7.9$ GeV are also taken from Ref.~\cite{Fu:2021wde}.   Taking into account isospin symmetry, the relevant couplings  are obtained as $g_{D_{s0}^{\ast+}D^{+}K^{0}}=g_{D_{s0}^{\ast+}D^{0}K^{+}}=\frac{1}{\sqrt{2}}g_{D_{s0}^*DK}$ and $g_{D_{s1}^+D^{\ast+}K^{0}}=g_{D_{s1}^{+}D^{\ast0}K^{+}}=\frac{1}{\sqrt{2}}g_{D_{s1}D^*K}$.

\subsection{Decay amplitudes and partial decay widths}

The decay amplitudes of $B^{+(0)}\to  D_{s}^{(\ast)+} \bar{D}^{(\ast)0}(D^{(\ast)-})$ and  $B^{+(0)}\to J/\psi(\eta_{c}) K^{+(0)}$ can be written as the products of two hadronic matrix elements~\cite{Ali:1998eb,Qin:2013tje}
\begin{eqnarray}\label{Ds-KK}
\mathcal{A}\left(B^{+}\to D_{s}^{+} \bar{D}^{\ast0}\right)&=&\frac{G_{F}}{\sqrt{2}} V_{cb}V_{cs} a_{1}\left\langle D_{s}^{+}|(s\bar{c})| 0\right\rangle\left\langle \bar{D}^{\ast 0}|(c \bar{b})| B^{+}\right\rangle , \\
\mathcal{A}\left(B^{+}\to D_{s}^{+} \bar{D}^{0}\right)&=&\frac{G_{F}}{\sqrt{2}} V_{cb}V_{cs} a_{1}^{\prime}\left\langle D_{s}^{+}|(s\bar{c})| 0\right\rangle\left\langle \bar{D}^{ 0}|(c \bar{b})| B^{+}\right\rangle , \\
\mathcal{A}\left(B^{+}\to D_{s}^{\ast+} \bar{D}^{0}\right)&=&\frac{G_{F}}{\sqrt{2}} V_{cb}V_{cs} a_{1}^*\left\langle D_{s}^{\ast+}|(s\bar{c})| 0\right\rangle\left\langle \bar{D}^{ 0}|(c \bar{b})| B^{+}\right\rangle , \\ 
\mathcal{A}\left(B^{+}\to D_{s}^{\ast+} \bar{D}^{\ast0}\right)&=&\frac{G_{F}}{\sqrt{2}} V_{cb}V_{cs} a_{1}^{\prime\ast}\left\langle D_{s}^{\ast+}|(s\bar{c})| 0\right\rangle\left\langle \bar{D}^{\ast 0}|(c \bar{b})| B^{+}\right\rangle , \\
\mathcal{A}\left(B^{+}\to J/\psi {K}^+\right)&=&\frac{G_{F}}{\sqrt{2}}V_{cb}V_{cs} a_{2}\left\langle J/\psi|(\bar{c} c)| 0\right\rangle\left\langle K^{+}|(s\bar{b} )| B^{+}\right\rangle,\\
\mathcal{A}\left(B^{+}\to \eta_{c} {K}^+\right)&=&\frac{G_{F}}{\sqrt{2}}V_{cb}V_{cs} a_{2}^{\prime}\left\langle \eta_{c}|(\bar{c} c)| 0\right\rangle\left\langle K^{+}|(s\bar{b} )| B^{+}\right\rangle,
\end{eqnarray}
where $a_{1}=c_{1}^{e f f}+c_{2}^{e f f} / N_{c}$ and $a_{2}=c_{1}^{e f f}/ N_{c}+c_{2}^{e f f} $ with $N_c$ the number of colors. It should be noted that $a_1$ and $a_{2}$ can be obtained in the factorization approach~\cite{Bauer:1986bm}.

 The current matrix elements between a pseudoscalar meson or vector meson and the vacuum have the following form:
\begin{eqnarray}
\left\langle D_{s}^{+}|(s \bar{c})| 0\right\rangle & =& f_{D_{s}^{+}} p^{\mu}_{D_{s}^{+}}, ~~
\left\langle D_{s}^{\ast+} |(s\bar{c} )| 0\right\rangle = m_{D_{s}^{\ast+}}f_{D_{s}^{\ast+}}\epsilon_\mu^*,  \\ \nonumber  ~~ \left\langle \eta_{c}|(c \bar{c})| 0\right\rangle& =& f_{\eta_{c}} p^{\mu}_{\eta_{c}}, ~~~~~~~
\left\langle J/\psi |(\bar{c} c)| 0\right\rangle = m_{J/\psi}f_{J/\psi}\epsilon_\mu^*,
\end{eqnarray}
where $f_{D_{s}^{+}}$, $f_{D_{s}^{\ast+}}$, $f_{\eta_{c}}$, and $f_{J/\psi}$ are the decay constants for $D_s^{+}$, $D_s^{\ast+}$, $\eta_{c}$, and $J/\psi$, respectively, and $\epsilon_\mu^*$ denotes the polarization vector of a vector particle.  In this work, we take  $G_F = 1.166 \times 10^{-5}~{\rm GeV}^{-2}$, $V_{cb}=0.041$, $V_{cs}=0.987$, $f_{D_{s}} = 250$ MeV, $f_{D_{s}^{\ast+}}=272$~MeV,  $f_{J/\psi} = 405$ MeV, and $f_{\eta_{c}} = 420$ MeV as in Refs.~\cite{ParticleDataGroup:2020ssz,Verma:2011yw,FlavourLatticeAveragingGroup:2019iem,Donald:2012ga,Li:2017mlw}.  
 
The hadronic matrix elements can be parameterised in terms of form factors~\cite{Verma:2011yw}
\begin{eqnarray}
&&\left\langle \bar{D}^{\ast0}|(c\bar{b})| B^{+}\right\rangle=\epsilon_{\alpha}^{*}\left\{-g^{\mu \alpha} (m_{\bar{D}^{\ast0}}+m_{B^{+}}) A_{1}\left(q^{2}\right)+P^{\mu} P^{\alpha} \frac{A_{2}\left(q^{2}\right)}{m_{\bar{D}^{\ast0}}+m_{B^{+}}}\right. \\  \nonumber
&&+i \varepsilon^{\mu \alpha \beta \gamma}P_\beta q_\gamma \left.\frac{V\left(q^{2}\right)}{m_{\bar{D}^{\ast0}}+m_{B^{+}}} +q^{\mu} P^{\alpha} \left[\frac{m_{\bar{D}^{\ast0}}+m_{B^{+}}}{q^{2}}A_{1}\left(q^{2}\right)-\frac{m_{B^{+}}-m_{\bar{D}^{\ast0}}}{q^{2}}A_{2}\left(q^{2}\right)-\frac{2m_{\bar{D}^{\ast0}}}{q^{2}}A_{0}\left(q^{2}\right)\right]\right\},
\\ 
&&\left\langle \bar{D}^{0}|(c \bar{b} )| B^{+}\right\rangle =\left[(p_{B^{+}}+p_{\bar{D}^{0}})^{\mu}-\frac{m_{B^+}^2-m_{\bar{D}^0}^2}{q'^2}q'_{\mu}\right] F_{1D}(q'^2)+\frac{m_{B^+}^2-m_{\bar{D}^0}^2}{q'^2}q'_{\mu} F_{0D}(q'^2),\\ 
&&\left\langle K^+|(s \bar{b} )| B^{+}\right\rangle =\left[(p_{B^{+}}+p_{K^{+}})^{\mu}-\frac{m_{B^+}^2-m_{K^{+}}^2}{q''^2}q''_{\mu}\right] F_{1K}(q''^2)+\frac{m_{B^+}^2-m_{K^+}^2}{q''^2}q''_{\mu} F_{0K}(q''^2),
\end{eqnarray}
where $q$,  $q^{\prime}$ and $q^{\prime\prime}$   represent the momentum transfer of $ p_{B^{+}}- p_{\bar{D}^{\ast0}}$, $ p_{B^{+}}- p_{\bar{D}^{0}}$, and $ p_{B^{+}}- p_{K^+}$, respectively, and $P = p_{B^{+}} + p_{\bar{D}^{\ast0}}$.  

The form factors of  $F_{1,0D}(t)$,  $F_{1,0K}(t)$, $A_{0}(t)$, $A_{1}(t)$, $A_{2}(t)$, and $V(t)$ with $t \equiv q^{\prime(\prime\prime)2}$ can be parameterized as~\cite{Verma:2011yw} 
\begin{equation}
X(t)=\frac{X(0)}{1-a\left(t / m_{B}^{2}\right)+b\left(t^{2} / m_{ B}^{4}\right)}.
\end{equation}
For these form factors, we adopt those of the covariant light-front quark model, i.e.,
  $(F_1(0), a, b)^{B \to \bar{D}} = (0.67, 1.22, 0.36)$,  $(F_0(0), a, b)^{B \to \bar{D}} = (0.67, 0.63, 0.01)$, $(F_1(0), a, b)^{B \to K }= (0.34, 1.60, 0.73)$, $(F_0(0), a,$ $b)^{B \to K} = (0.34, 0.78, 0.05)$, $(A_0(0), a, b)^{B \to \bar{D}^{\ast}} = (0.68, 1.21, 0.36)$, $(A_1(0), a, b)^{B \to \bar{D}^{\ast}} = (0.65, 0.60, 0.00)$,  $(A_2(0), a, b)^{B \to \bar{D}^{\ast}} = ( 0.61, 1.12, 0.31)$, and $(V_0(0), a, b)^{B \to \bar{D}^{\ast}} = ( 0.77, 1.25, 0.38)$~\cite{Verma:2011yw}.

With the above relevant Lagrangians, one can easily compute  the corresponding decay amplitudes of Fig.~\ref{triangle2317},  
\begin{align}\label{triangleA}
\mathcal{A}_{a} & = g_{D_{s0}^*D_{s}\eta}\int \frac{d^{4} q_{3}}{(2 \pi)^{4}} \frac{{\rm i}\mathcal{A}(B\to D_{s}\bar{D}^{\ast})\mathcal{A}\left(\bar{D}^{\ast} \to \bar{D} \eta\right)}{\left(q_{1}^{2}-m_{\bar{D}^{\ast}}^{2}\right)\left(q_{2}^{2}-m_{D_{s}}^{2}\right)\left(q_{3}^{2}-m_{\eta}^{2}\right)},  \\
\mathcal{A}_{b} & = g_{D_{s0}^*DK}\int \frac{d^{4} q_{3}}{(2 \pi)^{4}} \frac{{\rm i}\mathcal{A}(B\to J/\psi K)\mathcal{A}\left(J/\psi \to \bar{D} D\right)}{\left(q_{1}^{2}-m_{\psi}^{2}\right)\left(q_{2}^{2}-m_{K}^{2}\right)\left(q_{3}^{2}-m_{D}^{2}\right)}, \\
\mathcal{A}_{c} & = g_{D_{s0}^*D_{s}\eta}\int \frac{d^{4} q_{3}}{(2 \pi)^{4}} \frac{{\rm i}\mathcal{A}(B\to D_{s}\bar{D})\mathcal{A}\left(\bar{D}\to \bar{D}^{\ast} \eta\right)}{\left(q_{1}^{2}-m_{\bar{D}}^{2}\right)\left(q_{2}^{2}-m_{D_{s}}^{2}\right)\left(q_{3}^{2}-m_{\eta}^{2}\right)},  \\
\mathcal{A}_{d} & = g_{D_{s0}^*DK}\int \frac{d^{4} q_{3}}{(2 \pi)^{4}} \frac{{\rm i}\mathcal{A}(B\to \eta_{c} K)\mathcal{A}\left(\eta_{c} \to \bar{D}^{\ast} D\right)}{\left(q_{1}^{2}-m_{\eta_{c}}^{2}\right)\left(q_{2}^{2}-m_{K}^{2}\right)\left(q_{3}^{2}-m_{D}^{2}\right)}, 
\label{figa}
\end{align}
where $q_1$, $q_2$, and $q_3$  denote the momenta of $\bar{D}^{\ast}$, $D_{s}$, and $\eta$ for Fig.~\ref{triangle2317} (a), $J/\psi$, $K$, and $D$ for Fig.~\ref{triangle2317} (b), $\bar{D}$, $D_{s}$, and $\eta$ for Fig.~\ref{triangle2317} (c), and $\eta_{c}$, $K$, and $D$ for Fig.~\ref{triangle2317} (d), and  $p_{1}$ and $p_{2}$ represent the momenta of $\bar{D}^{(\ast)}$ and $D_{s0}^*(2317)$. 

Similarly, the corresponding decay amplitudes of Fig.~\ref{triangle24601} are written as 
\begin{align}\label{triangleb}
\mathcal{A}_{a} & = \int \frac{d^{4} q_{3}}{(2 \pi)^{4}} \frac{{\rm i}\mathcal{A}(B\to D_{s}^{\ast}\bar{D})\mathcal{A}\left(\bar{D}^{\ast} \to \bar{D} \eta\right)\mathcal{A}(D_{s}^*\eta\to D_{s1})}{\left(q_{1}^{2}-m_{\bar{D}^{\ast}}^{2}\right)\left(q_{2}^{2}-m_{D_{s}^*}^{2}\right)\left(q_{3}^{2}-m_{\eta}^{2}\right)},  \\
\mathcal{A}_{b} & = \int \frac{d^{4} q_{3}}{(2 \pi)^{4}} \frac{{\rm i}\mathcal{A}(B\to \eta_{c} K)\mathcal{A}\left(\eta_{c} \to \bar{D} D^*\right)\mathcal{A}(D^* K\to D_{s1})}{\left(q_{1}^{2}-m_{\psi}^{2}\right)\left(q_{2}^{2}-m_{K}^{2}\right)\left(q_{3}^{2}-m_{D^*}^{2}\right)},
\end{align}
and the  corresponding amplitudes of Fig.~\ref{triangle24602} are written as  
\begin{align}\label{trianglec}
\mathcal{A}_{a} & =\int \frac{d^{4} q_{3}}{(2 \pi)^{4}} \frac{{\rm i}\mathcal{A}(B\to D_{s}^{\ast}\bar{D}^{\ast})\mathcal{A}\left(\bar{D}^{\ast} \to \bar{D}^{\ast} \eta\right)\mathcal{A}(D_{s}^*\eta\to D_{s1})}{\left(q_{1}^{2}-m_{\bar{D}^{\ast}}^{2} \right)\left(q_{2}^{2}-m_{D_{s}^*}^{2}\right)\left(q_{3}^{2}-m_{\eta}^{2}\right)},  \\
\mathcal{A}_{b} & = \int \frac{d^{4} q_{3}}{(2 \pi)^{4}} \frac{{\rm i}\mathcal{A}(B\to D_{s}^*\bar{D})\mathcal{A}\left(\bar{D}\to \bar{D}^{\ast} \eta\right)\mathcal{A}(D_{s}^*\eta\to D_{s1})}{\left(q_{1}^{2}-m_{\bar{D}}^{2}\right)\left(q_{2}^{2}-m_{D_{s}}^{2}\right)\left(q_{3}^{2}-m_{\eta}^{2}\right)},\\
\mathcal{A}_{c} & = \int \frac{d^{4} q_{3}}{(2 \pi)^{4}} \frac{{\rm i}\mathcal{A}(B\to J/\psi K)\mathcal{A}\left(J/\psi \to \bar{D}^* D^*\right)\mathcal{A}(D^* K\to D_{s1})}{\left(q_{1}^{2}-m_{\psi}^{2} \right)\left(q_{2}^{2}-m_{K}^{2}\right)\left(q_{3}^{2}-m_{D^*}^{2}\right)},   \\
\mathcal{A}_{d} & =\int \frac{d^{4} q_{3}}{(2 \pi)^{4}} \frac{{\rm i}\mathcal{A}(B\to \eta_{c} K)\mathcal{A}\left(\eta_{c} \to \bar{D}^{\ast} D^*\right)\mathcal{A}(D^* K\to D_{s1})}{\left(q_{1}^{2}-m_{\eta_{c}}^{2}\right)\left(q_{2}^{2}-m_{K}^{2}\right)\left(q_{3}^{2}-m_{D^*}^{2}\right)}, 
\end{align}
where the representation of momenta  are the same as Eqs.~(16-19).

The  weak decay amplitudes of $B\to D_{s}^{(\ast)}\bar{D}^{(\ast)}$ and $B\to J/\psi(\eta_{c}) K$ are written as 
\begin{align}\label{am3}
\mathcal{A}(B\to D_{s}\bar{D}^{\ast})&= \frac{G_{F}}{\sqrt{2}}V_{cb}V_{cs} a_{1} f_{D_{s}}\{-q_{2}\cdot \varepsilon(q_{1})(m_{\bar{D}^{\ast0}}+m_{B^{+}}) A_{1}\left(q_{2}^{2}\right)   \\ \nonumber 
&+(k_{0}+q_{1}) \cdot \varepsilon(q_{1}) q_{2}\cdot (k_{0}+q_{1}) \frac{A_{2}\left(q_{2}^{2}\right)}{m_{\bar{D}^{\ast0}}+m_{B^{+}}} +(k_{0}+q_{1}) \cdot \varepsilon(q_{1}) \\ \nonumber       & 
[(m_{\bar{D}^{\ast0}}+m_{B^{+}})A_{1}(q_{2}^2) -(m_{B^{+}}-m_{\bar{D}^{\ast0}})A_{2}(q_2^2) -2m_{\bar{D}^{\ast0}} A_{0}(q_{2}^2)]  \} , \\ \nonumber
\mathcal{A}(B\to D_{s}\bar{D})&=\frac{G_{F}}{\sqrt{2}}V_{cb}V_{cs} a_{1}^{\prime}f_{D_{s}}(m_{B}^2-m_D^2)F_{0D}(q_{2}^2),
\\ \nonumber
\mathcal{A}(B^{+}\to D_{s}^{\ast+}\bar{D}^{0})&= \frac{G_{F}}{\sqrt{2}}V_{cb}V_{cs} a_{1}^{*}m_{D_{s}^{\ast+}}f_{D_{s}^{\ast+}}(k_{0}+q_{1})^{\mu}F_{1}(q_{2}^2), \\ \nonumber
 \mathcal{A}(B^{+}\to D_{s}^{\ast+}\bar{D}^{\ast0})&= \frac{G_{F}}{\sqrt{2}}V_{cb}V_{cs} a_{1}^{*\prime}m_{D_{s}^{\ast+}}f_{D_{s}^{\ast+}}\left[(-g^{\mu \alpha} (m_{\bar{D}^{\ast0}}+m_{B^{+}}) A_{1}\left(q_2^{2}\right)\right.  \\ \nonumber &+ \left. P^{\mu} P^{\alpha} \frac{A_{2}\left(q_2^{2}\right)}{m_{\bar{D}^{\ast0}}+m_{B^{+}}}  
+i \varepsilon^{\mu \alpha \beta \gamma}P_\beta q_\gamma \frac{V\left(q_2^{2}\right)}{m_{\bar{D}^{\ast0}}+m_{B^{+}}}\right],  \\ \nonumber
\mathcal{A}(B\to J/\psi K)&=\frac{G_{F}}{\sqrt{2}}V_{cb}V_{cs} a_{2} m_{\psi}f_{\psi}\varepsilon(q_{1})\cdot (k_{0}+q_{2}) F_{1K}(q_{1}^2),\\ \nonumber
\mathcal{A}(B\to \eta_{c} K)&=\frac{G_{F}}{\sqrt{2}}V_{cb}V_{cs} a_{2} f_{\eta_{c}}(m_{B}^2-m_K^2)F_{0K}(q_{1}^2).
\end{align}
With these branching ratios of $B^{+(0)}\to  D_{s}^{(\ast)+} \bar{D}^{(\ast)0}(D^{(\ast)-})$ and  $B^{+(0)}\to J/\psi(\eta_{c}) K^{+(0)}$ in Table~\ref{ratios},  we determine  $a_1=0.93(0.95)$,      $a_1^{\prime}=0.80(0.74)$,  $a_{1}^{ \ast}=0.81(0.83)$, and  $a_{1}^{\prime\ast}=0.83(0.88)$ as well as $a_{2}=0.27(0.26)$ and   $a_{2}^{\prime}=0.24(0.21)$,  consistent with the estimates of Ref.~\cite{Ali:1998eb}.

The vertices representing the $\bar{D}^{(\ast)}$ mesons scattering into $\bar{D}^{(\ast)}$ and $\eta$ mesons and $J/\psi(\eta_{c})$ mesons scattering into $\bar{D}^{(\ast)}$ and $D^{(\ast)}$ mesons are written as 
\begin{eqnarray}
\mathcal{A}\left(\bar{D}^{\ast} \to \bar{D} \eta\right)&=&g_{\bar{D}^{\ast}\bar{D}\eta}q_{3}\cdot \varepsilon(q_{1}),\\
\mathcal{A}\left(\bar{D} \to \bar{D}^{\ast} \eta\right)&=&-g_{\bar{D}^{\ast}\bar{D}\eta}q_{3}\cdot \varepsilon(q_{1}),  \\
  \mathcal{A}\left(\bar{D}^*\to \bar{D}^{\ast} \eta\right) &=& g_{\bar{D}^{\ast}\bar{D}^{\ast} \eta}\varepsilon_{\mu\nu\alpha\beta}q_{1}^{\mu}\varepsilon^{\nu}(q_{1})p_{1}^{\alpha}\varepsilon^{\beta}(p_{1}), \\
\mathcal{A}(J/\psi \to \bar{D} D)&=&-m_{\psi}/f_{\psi} (q_{3}-p_{1})\cdot \varepsilon(q_{1}), \\
\mathcal{A}(\eta_{c} \to \bar{D}^{\ast} D)&=&g_{\eta_{c} \bar{D}^* D}(q_{3}+q_{1})\cdot\varepsilon(p_{1}),\\
\mathcal{A}\left(J/\psi \to \bar{D}^{\ast} D^*\right) &=& g_{J/\psi\bar{D}^{\ast} D^*} \left[ \varepsilon(q_{1})^{\mu}(p_{1}-q_{3})_{\mu}\varepsilon(q_3)^{\nu}\varepsilon(p_1)_{\nu}+\varepsilon(p_1)^{\mu}(q_1+q_3)_{\mu}\varepsilon(q_1)^{\nu}\varepsilon(q_3)_{\nu} \right. \nonumber \\
&& \left. -\varepsilon(q_3)^{\mu}(p_1+q_1)_{\mu}\varepsilon(q_1)^{\nu}\varepsilon(p_1)_{\nu} \right], \\ 
\mathcal{A}\left(\eta_{c} \to \bar{D}^{\ast} D^*\right) &=& g_{\bar{D}^{\ast}\bar{D}^{\ast} \eta_{c}}\varepsilon_{\mu\nu\alpha\beta}q_{3}^{\mu}\varepsilon^{\nu}(q_{3})p_{1}^{\alpha}\varepsilon^{\beta}(p_{1}). 
\end{eqnarray}

The vertices describing the  $D_{s0}^*(2317)$ and $D_{s1}(2460)$ molecules generated by  $D^{(\ast)}K$ and $D_{s}^{(\ast)}\eta$ coupled channels are expressed as 
\begin{eqnarray}
\mathcal{A}(DK\to D_{s0}^*) &=& g_{D_{s0}^*DK}, \\
\mathcal{A}(D_{s}\eta\to D_{s0}^*) &=& g_{D_{s0}^*D_{s}\eta}, \\
\mathcal{A}(D_{s}^*\eta\to D_{s1}) &=& g_{D_{s}^*\eta D_{s1}}\varepsilon(p_{2})\cdot\varepsilon(q_3), \\
\mathcal{A}(D^* K\to D_{s1}) &=& g_{D^* K D_{s1}}\varepsilon(p_{2})\cdot\varepsilon(q_3).
\end{eqnarray}

 With the above amplitudes determined as specified above, 
 the corresponding partial decay widths  can be finally written as
 \begin{eqnarray}
\Gamma={8\pi}\frac{|\vec{p}\,|}{m_{B}^2}\overline{|\mathcal{M}|}^{2},
\end{eqnarray}
where the overline indicates the sum over the polarization vectors of final states, and $|\vec{p}\,|$ is the momentum of either final state in the rest frame of  the $B$ meson.

\section{Numerical Results and Discussion}\label{sec:Results}

\begin{table}[ttt]
\caption{Masses and quantum numbers of  mesons relevant to the present work~\cite{ParticleDataGroup:2020ssz}. \label{mass}}
\begin{tabular}{ccc|ccc}
  \hline\hline
   Meson & $I (J^P)$ & M (MeV) &    Meson & $I (J^P)$ & M (MeV)   \\
  \hline
     $B^{0}$ & $\frac{1}{2}(0^-)$ & $5279.65$  &    $B^{+}$ & $\frac{1}{2}(0^-)$ & $5279.34$ \\
   $D^{0}$ & $\frac{1}{2}(0^-)$ & $1864.84$  &    $D^{+}$ & $\frac{1}{2}(0^-)$ & $1869.66$ \\
  $D^{\ast0}$ & $\frac{1}{2}(1^-)$ & $2006.85$ &  $D^{\ast+}$ & $\frac{1}{2}(1^-)$ & $2010.26$   \\
      $D_{s}^{+}$ & $0(0^-)$ & $1968.34$ & 
  $D_{s}^{\ast+}$ & $0(1^-)$ & $2112.2$ \\
  $D_{s0}^*$ & $0(0^+)$ & $2317.8$ & 
  $D_{s1}$ & $0(1^+)$ & $2459.5$ \\
  $K^{+}$ & $\frac{1}{2}(0^-)$ & $493.677$ & 
  $K^{0}$ & $\frac{1}{2}(0^-)$ & $497.611$ \\
    $\eta$ & $0(0^-)$ & $547.862$ & 
  $J/\psi$ & $0(1^-)$ & $3096.9$ \\
 \hline \hline
\end{tabular}
\label{tab:masses}
\end{table}

With the above preparation and the masses of relevant particles given in Table~\ref{mass}, we can obtain the decay widths of  $B \to \bar{D}^{(\ast)}D_{s0}^{*}(2317) $ shown in Table~\ref{results1}. We note that the branching ratios of  $B^{+}\to \bar{D}^{0}D_{s0}^{\ast+}(2317) $, $B^{0}\to {D}^{-}D_{s0}^{\ast+}(2317) $, $B^{+}\to \bar{D}^{\ast0}D_{s0}^{\ast+}(2317) $,  and $B^{0}\to {D}^{\ast-}D_{s0}^{\ast+}(2317)$  are consistent with the experimental data within uncertainties~\cite{ParticleDataGroup:2020ssz} .   The theoretical uncertainties originate from the breaking of $SU(3)$-flavor symmetry and heavy quark spin symmetry, which are used in  deriving the couplings of $g_{\bar{D}^{\ast}\bar{D}\eta}$ and $g_{\eta_{c}/J/\psi \bar{D}^{(\ast)}D}$. We assume that the breaking of $SU(3)$-flavor symmetry is at the order of   $20\%$ and  that of heavy quark spin symmetry is at the level of $20\%$~\cite{Liu:2019stu}. Adding them in quadrature, we obtain the theoretical uncertainty of $28\%$ given in Table~\ref{results}.

\begin{table}[!h]
\centering
\caption{Branching ratios ($10^{-3}$) of $B\to \bar{D}^{(\ast)}D_{s0}^*(2317)$ and $B\to \bar{D}^{(\ast)}D_{s1}(2460)$. \label{results}
}
\label{results1}
\begin{tabular}{c c c c c c c c c}
  \hline \hline
    decay modes    &~~~~ Our results  &~~~~ \cite{Faessler:2007cu}    &~~~~ RPP~\cite{ParticleDataGroup:2020ssz}  &~~~~ BarBar
         \\ \hline 
        $B^{+} \to \bar{D}^{0}D_{s0}^{*+}(2317) $   &~~~~ $0.677\pm 0.190$   &~~~~ $1.03\pm 0.14$ &~~~~ $0.80^{+0.16}_{-0.13}$    & ~~~~ $1.0\pm 0.3 \pm 0.1$  
                 \\   $B^{0} \to D^{-}D_{s0}^{*+}(2317) $   &~~~~ $0.637\pm0.178 $  &~~~~ $0.96\pm 0.13$  &~~~~ $1.06^{+0.16}_{-0.16}$  & ~~~~ $1.8\pm 0.4 \pm 0.3$  
    
         \\   $B^{+} \to \bar{D}^{\ast0}D_{s0}^{*+}(2317) $  &~~~~ $1.210\pm0.339$    &~~~~$0.50\pm 0.07$ &~~~~ $0.90^{+0.70}_{-0.70}$ & ~~~~ $0.9\pm 0.6 \pm 0.2$   
         \\     $B^{0} \to D^{\ast-}D_{s0}^{*+}(2317) $    &~~~~ $0.889\pm0.249$   &~~~~ $0.47\pm 0.06$  &~~~~ $1.50^{+0.60}_{-0.60}$  & ~~~~ $1.5\pm 0.4 \pm 0.2$  
         
             \\\hline
                $B^{+} \to \bar{D}^{0}D_{s1}^{+}(2460) $   &~~~~ $1.255\pm{0.351}$   &~~~~ $2.54\pm 0.39$ &~~~~ $3.1^{+1.0}_{-0.9}$ & ~~~~ $2.7\pm 0.7 \pm 0.5$
                 \\   $B^{0} \to D^{-}D_{s1}^{+}(2460) $   &~~~~ $1.158\pm 0.324$  &~~~~ $2.36\pm 0.36$  &~~~~ $3.5\pm1.1$ & ~~~~ $2.8\pm 0.8 \pm 0.5$ 
         \\                $B^{+} \to \bar{D}^{\ast0}D_{s1}^{+}(2460) $   &~~~~ $3.065\pm 0.858$   &~~~~ $7.33\pm 1.12$ &~~~~ $12.0\pm3.0$  & ~~~~ $7.6\pm 1.7 \pm 1.8$  
                 \\   $B^{0} \to D^{\ast-}D_{s1}^{+}(2460) $   &~~~~ $2.709 \pm 0.759$  &~~~~ $6.85\pm 1.05$  &~~~~ $9.3\pm2.2$ & ~~~~ $5.5\pm 1.2 \pm 1.0$   
         \\  
  \hline \hline
\end{tabular}
\end{table}

In Ref.~\cite{Faessler:2007cu}, the authors estimated the branching ratios of $B\to \bar{D}^{(\ast)}D_{s0}^*(2317)$ by the naive factorisation approach, where the coupling $f_{D_{s0}^*}$ is determined treating $D_{s0}^*(2317)$ as a pure $DK$ molecule. Their  branching ratios are shown in Table~\ref{results1}.  We note that the branching ratios of $B \to \bar{D} D_{s0}^*(2317)$ and $B \to \bar{D}^{\ast} D_{s0}^*(2317)$ are consistent with ours, but those of $B^{+} \to \bar{D}^{*0} D_{s0}^*(2317)$ and $B^{0} \to \bar{D}^{\ast-} D_{s0}^*(2317)$ are smaller than ours and in worse agreement with the experimental data. We note that many recent works claim that $D_{s0}^*(2317)$ contains a $c\bar{s}$ component of 30\%, which is not explicitly taken into account in both our work and Ref.~\cite{Faessler:2007cu}. Considering such an uncertainty, both our results and those of Ref.~\cite{Faessler:2007cu} are consistent with the experimental data.

\begin{table}[!h]
\centering
\caption{Branching ratios ($10^{-3}$) of $B\to \bar{D}^{(\ast)}D_{s0}^*(2317)$ and $B\to \bar{D}^{(\ast)}D_{s1}(2460)$. \label{resultsd}
}
\begin{tabular}{c c c c c c c c c}
  \hline \hline
    decay modes    &~~~~ Total results  &~~~~ $\eta$ meson exchange    &~~~~$D^{(\ast)}$ meson exchange 
         \\ \hline 
        $B^{+} \to \bar{D}^{0}D_{s0}^{*+}(2317) $   &~~~~ $0.677 $   &~~~~ $0.414$ &~~~~ $0.033$     
                 \\   $B^{0} \to D^{-}D_{s0}^{*+}(2317) $   &~~~~ $0.637  $  &~~~~ $0.401$  &~~~~ $0.028$  
    
         \\   $B^{+} \to \bar{D}^{\ast0}D_{s0}^{*+}(2317) $  &~~~~ $1.210 $    &~~~~$0.246$ &~~~~ $0.382$   
         \\     $B^{0} \to D^{\ast-}D_{s0}^{*+}(2317) $    &~~~~ $0.889 $   &~~~~ $0.194$  &~~~~ $0.264$   
         
             \\\hline
                $B^{+} \to \bar{D}^{0}D_{s1}^{+}(2460) $   &~~~~ $1.255 $   &~~~~ $0.209$ &~~~~ $0.442$ 
                 \\   $B^{0} \to D^{-}D_{s1}^{+}(2460) $   &~~~~ $1.158 $  &~~~~ $0.202$   &~~~~ $0.309$  
         \\                $B^{+} \to \bar{D}^{\ast0}D_{s1}^{+}(2460) $   &~~~~ $3.065 $   &~~~~ $1.263$ &~~~~ $0.648$   
                 \\   $B^{0} \to D^{\ast-}D_{s1}^{+}(2460) $   &~~~~ $2.709$  &~~~~ $1.298$ &~~~~ $0.446$   
         \\  
  \hline \hline
\end{tabular}
\end{table}

For the $D_{s1}(2460)$ state, our predictions for all the four processes studied are smaller than the PDG averages by about a factor of 3 and than the BaBar results by roughly a factor of 2. On the other hand, the results of Ref.~\cite{Faessler:2007cu} are in better agreement with the data.     {   In Ref.~\cite{Faessler:2007cu}, the authors estimated such branching ratios via a naive factorisation approach, where the  determination of the  couplings $f_{D_{s0}^*}$ and $f_{D_{s1}}$ depends on the choice of cutoff parameter  and relies on   the SU(4) symmetry which relates the weak vertices $D^{\ast}\to K W$ and $D\to K^{\ast} W$. Furthermore, in Ref.~[55], $D_{s0}^*(2317)$ and $D_{s1}(2460)$ are treated as pure $DK$ and $D^*K$ molecules, while in our approach it is shown that the coupled channel $D^{(*)}\eta$  plays an important role as well.} The discrepancy between our results and the experimental data can be attributed to either missing reaction mechanisms or the neglect of the likely existence of a relatively large $c\bar{s}$ component in the wave function of $D_{s1}(2460)$. In most of the unquenched quark models, both $D_{s1}(2460)$ and $D_{s0}^*(2317)$ contain sizable $c\bar{s}$ components, while the former contains a larger $c\bar{s}$ component. In addition, in the molecular picture, other reaction mechanisms than the triangle mechanism studied here can also contribute to the production of $D_{s0}^*(2317)$ and $D_{s1}(2460)$ in $B$ decays, such as those studied in Refs.~\cite{Sun:2015uva,Albaladejo:2016hae}.

 {     We decompose the contributions of the $\eta$ and $D^{(*)}$ exchanges in Table~\ref{resultsd}. Note that the processes mediated by the $\eta$ meson   contain  stronger weak-interaction vertices  but weaker strong-interaction scattering vertices with respect to those mediated by the $D^{(\ast)}$ meson, while the couplings of the   $D_{s0}^{*}(2317)$ and $D_{s1}(2460)$ molecules   to their constituents  $D^{(\ast)}K$ and $D_{s}^{(\ast)}\eta$   are approximately the same in the particle basis, i.e., $g_{D_{s0}^{*}D^{+}K^{0}}
\approx g_{D_{s0}^{*}D_{s}^{+}\eta}$ $(g_{D_{s1}D^{\ast+}K^{0}}
\approx g_{D_{s}D_{s}^{\ast+}\eta})$. From Table ~\ref{resultsd}, one can see that among the eight branching ratios studied, the contribution of  the $\eta$  exchange  is comparable to that of  the $D^{(\ast)}$ exchange  except for the processes $B\to \bar{D}D_{s0}^*(2317)$, where the $D^{(*)}$ contribution is accidentally one order of magnitude smaller that of the $\eta$ exchange.  }

\section{Summary and Discussion}
\label{sum}

To distinguish the nature of $D_{s0}^{\ast}(2317)$ as either a $DK$ molecule, a $c\bar{s}$ state, or a combination of both has motivated a lot of experimental and theoretical studies.  In this work, 
 we utilized the triangle mechanism to describe the decays of $B \to \bar{D}D_{s0}^*(2317) $ and $B \to \bar{D}^{\ast}D_{s0}^*(2317) $,  assuming  that  the $B$ meson  first weakly decays into $\bar{D}^{\ast}D_{s}$ and  $J/\psi K$, then $\bar{D}^{\ast}$ and $J/\psi$ mesons scatter to  $\bar{D}^{(\ast)}\eta$ and $\bar{D}^{\ast}D$, and finally  $D_{s0}^{\ast}(2317)$  is dynamically generated by the $DK$ and $D_{s}\eta$ coupled-channel interactions. Without any unknown parameters, we take the effective Lagrangian approach to calculate  the branching ratios as Br$[B^{+} \to \bar{D}^{0}D_{s0}^{*+}(2317)] =0.677\times 10^{-3}$ (Br$[B^{0} \to  {D}^{-}D_{s0}^{*+}(2317) ]=0.676\times 10^{-3}$), and Br$[B^{+} \to \bar{D}^{\ast0}D_{s0}^{*+}(2317)]=1.210\times 10^{-3}$ (Br$[B^{0} \to  {D}^{\ast-}D_{s0}^{*+}(2317)]=0.889\times 10^{-3}$), which are in reasonable agreement with the experimental data.

In the same  approach, we also investigated  the  decays of $B \to \bar{D}D_{s1}(2460) $ and $B \to \bar{D}^{\ast}D_{s1}(2460) $, where $D_{s1}(2460)$ is  dynamically generated by the $D^*K$ and $D_{s}^*\eta$ coupled-channel interactions. Our results,  Br$[B^{+} \to \bar{D}^{0}D_{s1}^+(2460)]  =~1.255\times 10^{-3}$ (Br$[B^{0} \to  {D}^{-}D_{s1}^{+}(460) ]=1.158\times 10^{-3}$), and Br$[B^{+} \to \bar{D}^{\ast0}D_{s1}^{+}(2460)]=3.065\times 10^{-3}$ (Br$[B^{0} \to  {D}^{\ast-}D_{s1}^{+}(2460)]=2.709\times 10^{-3}$), are smaller than the experimental central values by almost a factor of $2\sim 3$.  
 Such a deviation can be  attributed to either a smaller molecular component in the $D_{s1}(2460)$ wave function or reaction  mechanisms missing in the present work.       
 
 We note that the degree of agreement between our predictions and the experimental data  indeed provides further support for the molecular nature of $D_{s0}^*(2317)$ and $D_{s1}(2460)$. However, more precise data and further theoretical studies are needed in order to pin down the precise percentage of the $c\bar{s}$ and $D^{*}K$/$D_s^{(*)}\eta$ components in their wave functions.

\section*{Acknowledgments}

We thank  Valery E. Lyubovitskij for helpful discussions. 
This work is partly supported by the National Natural Science Foundation of China under Grants No. 12105007, No. 11735003, No. 11975041, No. 11961141004, No. 11961141012, No. 12075288, and No. 1210050997, and the fundamental Research Funds for the Central Universities, the Youth Innovation Promotion Association CAS,  the Project of Youth Backbone Teachers of Colleges and Universities of Henan Province (2020GGJS017),  the Youth Talent Support Project of Henan (2021HYTP002), the Open Project of Guangxi Key Laboratory of Nuclear Physics and Nuclear Technology, No.NLK2021-08, and China Postdoctoral
Science Foundation under Grants No. 2022M710317, and No. 2022T150036.

\bibliography{reference}
\end{document}